\documentclass[preprint,prb,aps,endfloats]{revtex4}
\usepackage{graphicx}
\usepackage{dcolumn}
\usepackage{amsmath,amssymb}
\usepackage{boxedminipage}
\usepackage{array}
\usepackage{flafter}
\usepackage[mathscr]{euscript}
\usepackage[tight,TABTOPCAP]{subfigure}
\usepackage[normalsize]{caption}
\usepackage{verbatim}
\newcolumntype{.}{D{.}{.}{1}}
\begin{document}
\title{Nano-indentation of a room-temperature ionic liquid film on silica: a computational experiment}
\author{P. Ballone$^{(1)}$, M. G. Del P{\'o}polo$^{(1)}$, S. Bovio$^{(2,3)}$, A. Podest{\`a}$^{(2,3)}$, P. Milani$^{(2,3)}$, and N. Manini$^{(3)}$}
\affiliation{(1) Atomistic Simulation Centre, Queen's University Belfast, Belfast BT7 1NN, UK}
\affiliation{(2) C.I.Ma.I.Na, Universit{\`a} degli Studi di Milano, via Celoria 16, 20133, Milano, Italy}
\affiliation{(3) Dipartimento di Fisica, Universit\`a degli Studi di Milano, Via Celoria 16, 20133 Milano, Italy}

\vskip 1.0 truecm

\date{\today}
\begin{abstract}
We investigate the structure of the [bmim][Tf$_2$N]/silica interface by simulating the indentation of
a thin ($4$~nm) [bmim][Tf$_2$N] film by a hard nanometric tip. The ionic liquid/silica interface is represented in
atomistic detail, while the tip is modelled by a spherical mesoscopic particle interacting via an effective short-range 
potential. Plots of the normal force ($F_z$) on the tip as a function of its distance from the silica surface highlight 
the effect of weak layering in the ionic liquid structure, as well as the progressive loss of fluidity in approaching the 
silica surface. The simulation results for $F_z$ are in near-quantitative agreement with new AFM data measured on the 
same [bmim][Tf$_2$N]/silica interface at comparable thermodynamic conditions.
\end{abstract}
\maketitle
\newpage
\section{Introduction}
\label{intro}
Room-temperature ionic liquids (RTILs\cite{welton}) represent a vast class of organic ionic systems actively investigated for
their potential role as versatile and environmentally friendly solvents\cite{solvents}. In recent years, several other applications have been
proposed, most notably in catalysis\cite{cath}, lubrication\cite{lubri,naina}, electrochemistry\cite{electrochem}, and photovoltaic 
power generation\cite{photo}.
The successful development of these applications, however, depends on the properties of RTIL interfaces with solid surfaces, which
are still poorly understood. To be precise, several experimental and a few computational studies have been carried out
to characterise these interfaces, but, because of their complexity and of the huge number of different RTIL and solid surface choices, the combined 
effort of several experimental and computational groups has until now at most scratched the surface of this gargantuan research field.

Sum frequency generation spectroscopy\cite{sum}, X-ray\cite{xray} and neutron\cite{neutro} scattering are among the most
relevant experimental techniques that have been used to characterise the structure of RTIL/solid interfaces. Most
experiments have been carried out for RTILs in contact with silica, mica and graphite, mainly because of the relatively inert character 
of these solid substrates, but measurements on TiO$_2$, noble- and near-noble metal surfaces have been carried out as well. 
Almost without exception, the results for samples close to room temperature reveal strong molecular layering close to the surface.

Mechanical, and, in particular, rheological properties, have been investigated using the surface force apparatus (SFA), probing the RTIL/mica
interface\cite{perkin1, perkin2, ueno} down to nanometric length scales. The results for very thin RTIL films in between mica layers point, 
first of all, to a very low shear viscosity of RTILs at the interface, thus emphasising the interest of ionic fluids as lubricants. 
Moreover, the measurement of the normal force $F_z$ as a function of the distance $z$ separating the two approaching surfaces in the SFA, 
displays apparent oscillations, consistently with the RTIL layering seen in structural experiments. 
The periodicity, amplitude and phase of the oscillations strictly depend on the RTIL under investigation, reflecting the competing role
of Coulomb forces and excluded volume effects\cite{perkin1, perkin2}.

Atomic force microscopy (AFM) could represent a decisive tool to characterise RTIL/solid interfaces, able, in principle, to provide a direct view of 
the structure, and to map local mechanical properties at the nm scale. Early AFM measurements\cite{naina, naina2} for RTILs of the imidazolium, 
pyrrolidimium and pyridimium families deposited as thin films on the oxidised silicon surface revealed a surprising variety of different behaviours,
depending on the surface hydroxylation and charge, on the water content, on the electric bias of the tip, and on the velocity of the surface scanning.
More importantly, these measurements provided evidence of solid-like behaviour for RTILs in confined geometries, even at conditions such
that the bulk equilibrium phase is the liquid. This observation, in turn, might explain the remarkable stability of ionogels made of inorganic
nanoparticles solvated into RTILs, that could find applications in electrochemical and photovoltaic devices\cite{graetzel}.

Recently, the debate on structural, dynamical and mechanical properties of RTIL/solid interfaces has been revived by new
AFM measurements carried out by two groups, investigating different but related systems. In the first case\cite{atkin},  the interface
separating a solid support from a thick, bulk-like RTIL slab has been considered, while in the second case\cite{milani1, milani2} thin 
sub-micron RTIL films on the same or similar solid surfaces have been prepared and characterised.

In Ref.~\onlinecite{atkin}, the solvation force ($F_z$) on a $20$~nm
AFM tip immersed in a thick, bulk-like  RTIL has been measured
as a function of distance from a solid surface for a variety of ammonium- and imidazolium-based RTILs supported on mica, silica, graphite and gold.
Room temperature data for mica display a sequence of regular jumps, each corresponding to the puncturing of a discrete 
layer, whose separation is comparable to the diameter of the molecular ions. 
The sharp discontinuities in the force at regularly spaced distances from the surface do not conform to the idea of a fluid film, suggesting
instead solid-like ordering of the RTIL in proximity of the solid support. As expected, the precise details of the force versus distance relation
depend somewhat on the RTIL. More important is the dependence on the substrate, the strong layering apparent in the case of mica being
progressively attenuated in going to silica, graphite and gold. On all substrates, only a few layers (five or six at most) can be identified by 
steps in the force.
This picture is consistent with the X-ray data of Ref.~\onlinecite{xray} (b).

Topographic AFM images of [bmim][Tf$_2$N] nanometric droplets on mica, silica and graphite have been reported in Ref.~\onlinecite{milani1}.
Measurements have been carried out in tapping mode, reaching a contact pressure under the tip ($\sim 0.2$ kbar) presumably
lower than that of the force-distance
measurements of Ref.~\onlinecite{atkin}. 
In the case of mica and silica, the AFM images reveal layered structures, consisting of the stacking of up to 100 RTIL planes, raising
more than 50~nm above the substrate terraces. The measured local mechanical properties and the long persistence time of the observed 
structures are fully consistent with the properties of solid-like RTIL islands. 
Even though solid-like layering of liquids wetting solid surfaces has been seen at several other interfaces,
the wide extension (up to 100 planes) of the solid-like structures is indeed unusual, taking into account that in the bulk the deposited compound
is liquid at the ambient temperature and pressure of experiment. The solid-like character, therefore, has to result from the in-plane
perturbation due to the substrate, which, however, is localised 50~nm below the topmost solid-like RTIL layers.
Also in these measurements, layering is greatly attenuated in the case of graphite and gold. 
The results of Ref.~\onlinecite{milani1} are qualitatively consistent with those reported in Ref.~\onlinecite{others}.

Similarities in the results of Ref.~\onlinecite{atkin} and Ref.~\onlinecite{milani1} are apparent in the strong layering observed in both 
experiments for mica, and in the common trends seen when going from mica to silica, graphite and gold.
Differences are present as well, since the layering seen in
Ref.~\onlinecite{milani1} extends far beyond the few molecular planes
observed in Ref.~\onlinecite{atkin}.

Strict comparison of the results of Ref.~\onlinecite{milani1} and Ref.~\onlinecite{atkin}, however, is prevented by the difference in the 
RTIL under consideration, and by the different system configuration (i.e., a thick, bulk-like film in Ref.~\onlinecite{atkin}, a distribution 
of nanometric droplets in Ref.~\onlinecite{milani1}) adopted in the two experiments. More importantly, the two sets of experiments do not
measure the same quantities, since Ref.~\onlinecite{atkin} reports primarily the normal force $F_z$ on the AFM tip immersed in the RTIL,
and approaching the solid substrate, while Ref.~\onlinecite{milani1} focuses on the topography of thin RTIL films. Moreover, the pressure
under the AFM tip is expected to reach a significantly higher value in the force-distance measurements of Ref.~\onlinecite{atkin} than
in the topographic imaging of Ref.~\onlinecite{milani1}. 

New AFM experiments\cite{milani2}, briefly discussed in Sec.~\ref{expe}, clarify the connection between the data of Ref.~\onlinecite{milani1} and 
Ref.~\onlinecite{atkin}, and show that the mechanical response of thin film is significantly different from that of contact layers in thick, 
bulk like RTIL slabs. This result, together with the computational study detailed in the next two sections, provide support and
insight on the unusual stability of solid-like [bmim][Tf$_2$N] islands on insulating, solid oxide surfaces.

\section{Model and simulation method}
\label{method}
To investigate local mechanical properties of thin RTIL films deposited on solid surfaces, we resort to computational approaches, 
simulating the indentation process of a nanometric film by a hard AFM tip, represented as a rigid sphere of nanometric diameter. 

In the present study we focus on the [bmim][Tf$_2$N]/silica interface, i.e., one of the systems considered in Ref.~\onlinecite{milani1}.
This same interface has been investigated by atomistic simulations\cite{wippf, us}, modelling either a bulk liquid/solid  termination, or 
a thin [bmim][Tf$_2$N] film deposited on silica, exposing a free surface to vacuum. This last configuration, in particular, was considered
in a series of computations carried out in our group\cite{us}, simulating a $\sim 4$~nm film on silica at temperatures $300 \leq T \leq 400$~K.

\begin{figure}[tbp]
\includegraphics[scale=1.0]{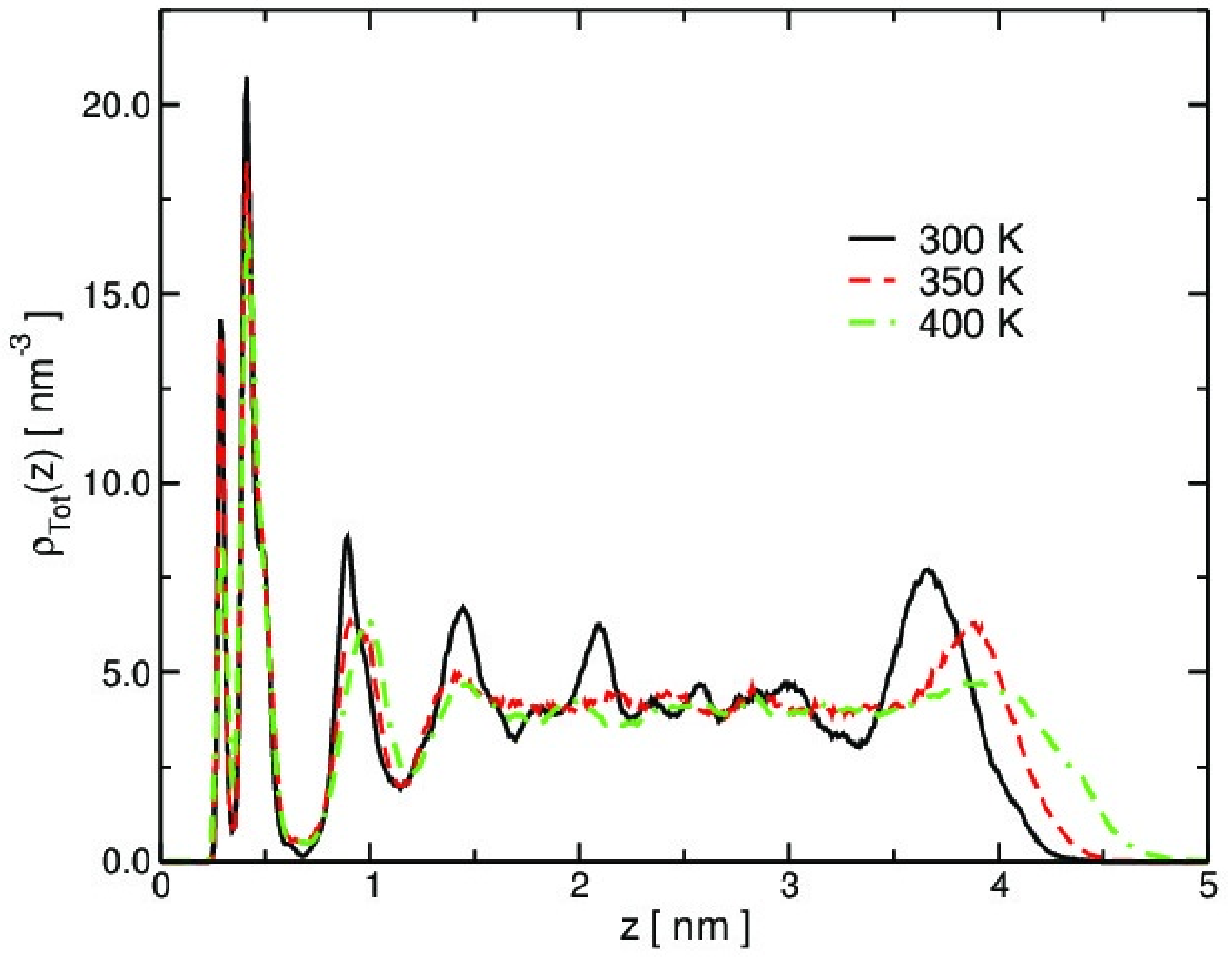}
\caption{Ion density profile as a function of distance from the silica surface.
Ions are represented as point particles, located at their centre of charge (see Ref.~\onlinecite{us}).
}
\label{rhoofz}
\end{figure}

A plot of the [bmim][Tf$_2$N] density profile in the direction perpendicular to the surface is shown in Fig.~\ref{rhoofz}. Layering is apparent
at the interface, but its amplitude decreases quickly in moving away from silica. Another remarkable feature is the density peak at
the [bmim][Tf$_2$N] free surface, due to the pile-up of cations lowering the surface energy and increasing entropy by exposing their
alkane tails towards the vacuum side of the interface.

Analysis of the ion mobility\cite{us} shows that diffusion is slow at all temperatures, and practically vanishes on the simulation time scale 
($\sim 10$~ns) in the first [bmim][Tf$_2$N] ionic layers. 
Ion diffusion is negligible across the entire film at $T=300$~K, a temperature at which at least four 
[bmim][Tf$_2$N]
layers
are clearly identified at the interface. At this temperature, the decay of the density oscillations away from
silica appears to be due more to the defective stacking of solid-like planes than to a genuine fluid-like dynamics of the ions. In other terms,
at least up to the film thickness ($\sim 4$~nm) and time scale ($\sim 10$~ns) covered by our simulations, the properties of the [bmim][Tf$_2$N] 
film appear to be more glassy than liquid-like.

Despite the detailed description provided by the atomistic simulations of Ref.~\onlinecite{us}, it is difficult to infer from these results whether 
the film will appear solid-like or fluid-like under probing by an AFM tip. To fill this gap, we carried out a molecular dynamics (MD) simulation 
of a nanometric tip penetrating into a thin [bmim][Tf$_2$N] film on silica. We adopted simulation protocols similar to those used by other 
authors\cite{ruth} for simpler solid/liquid interfaces.

The atomistic model of the RTIL film deposited on silica is described in detail in Refs.~\onlinecite{us, pote}. The validity,
reliability and accuracy of the interatomic potentials for [bmim][Tf$_2$N],
that represents the main
portion of 
our system, are extensively discussed in Ref.~\onlinecite{pote}. 

Experimental tips typically are made of a ceramic material (Si$_3$N$_4$ in the experiments of Ref.~\onlinecite{milani1}), with a radius
of $\sim 20-50$~nm, corresponding to a few million atoms. In general, the
tip carries a charge, that
could be controlled
by a suitable external
bias. In our computations we consider smaller, neutral tips. The simulated tip size, in particular, will range from $1.6$~nm to $2.0$~nm, corresponding 
to $\sim 1000-3000$ atoms. Within this range, the atomistic structure of the tip could affect the results for the normal 
force felt by the tip in proximity of the surface. However, we aim at comparing our data to experimental results obtained for much larger tips
($\sim 20$~nm), whose atomistic structure is likely to have only a minor effect. For this reason, we adopt a very idealised model of
the tip, while we briefly discuss the possible role of its electrostatic
charge in Sect.~\ref{expe}.

In what follows, the tip is represented by a spherical particle of infinite mass (nano-sphere, NS) interacting with all real atoms via a 
short-range potential. The radial dependence is given by:
\begin{equation}
V(r)=\frac{4 \epsilon}{\alpha^2} \left\{ \frac{1}{\left[ \left( \frac{r}{\sigma}\right)^2-1\right]^6}
-\alpha \frac{1}{\left[ \left( \frac{r}{\sigma}\right)^2-1\right]^3} \right\} \ \ \ \ \ \ \ \ \ \ \ \quad r > \sigma
\label{pote}
\end{equation}
This functional form of $V(r)$ has been proposed in Ref.~\onlinecite{daan} to model colloids and nanoparticles.
The parameters chosen for our simulation ($\sigma=1.6-2$~nm, $\epsilon=0.7$~kJ/mol, $\alpha=10$) are such to
describe a nanometric excluded volume, with a thin (sub-nanometric) attractive well outside the sphere. It is apparent from eq.~(\ref{pote})
that no atom can enter the sphere of radius $\sigma$ centred on the NP. Since atoms are $\sim 1000$ times smaller (in volume) than the
smallest NP we considered, it is easy to verify that the length $\sigma$ represents in this case the {\it radius} of the NP.
This is at variance from the case of equi-sizes LJ particles, for which the corresponding $\sigma_{LJ}$ parameter represents the
particle's diameter. This consideration will be relevant when estimating the pressure under the tip during the indentation process.

\section{Simulation results}
\label{results}
MD simulations have been carried out using a version of the DL$\_$POLY package\cite{DLPOLY}
slightly modified to account for the potential of eq.~(\ref{pote}).

The simulations of Ref.~\onlinecite{us} provided well equilibrated starting configurations for the present study. The simulated surface,
lying parallel to the $xy$ plane, is nearly square, with a cross section of $L_x \times L_y=5.5 \times 5.3$~nm$^2$. The solid side of the interface is 
represented by one oxygen plane and one silicon plane, whose atoms reproduce the geometry of the Si-terminated (111) surface of $\beta$-cristobalite. Each 
silicon atom
carries an OH group, corresponding to a coverage of $4.1$ OH/nm$^2$. All Si and O atoms in the system 
are kept fixed, while the H atoms of the hydroxyl terminations are mobile. The RTIL film is represented by $250$ [bmim]$^+$, [Tf$_2$N]$^-$ 
ion pairs, corresponding to a thickness of $4.5$~nm at $T=300$~K. The basic simulation box is replicated in 3D with periodicity 
$L_x \times L_y \times L_z=5.5 \times 5.3 \times 17$~nm$^3$, sufficient to accommodate a mesoscopic spherical tip interacting with the free surface of 
the RTIL, and still well separated from the periodic image of the silica surface. In what follows, the position of the meso-particle is identified by
the $z$ coordinate ($Z$) of its centre with respect to the topmost Si plane in the SiO$_2$ slab. The in-plane periodicity of the (111) cristobalite surface
is much shorter than the NS diameter, and therefore the precise location of the contact point in the $xy$ plane 
is relatively unimportant. Long-range Coulomb interactions are accounted for using 3D Ewald summation, with a relative convergence parameter
of $10^{-5}$. The NVT conditions of our simulations are enforced by a Nos{\'e}-Hoover thermostat\cite{DLPOLY}. The equations of motion are
integrated using the velocity Verlet algorithm with a time step of $1$ fs.

\begin{figure}[tbp]
\includegraphics[scale=0.3]{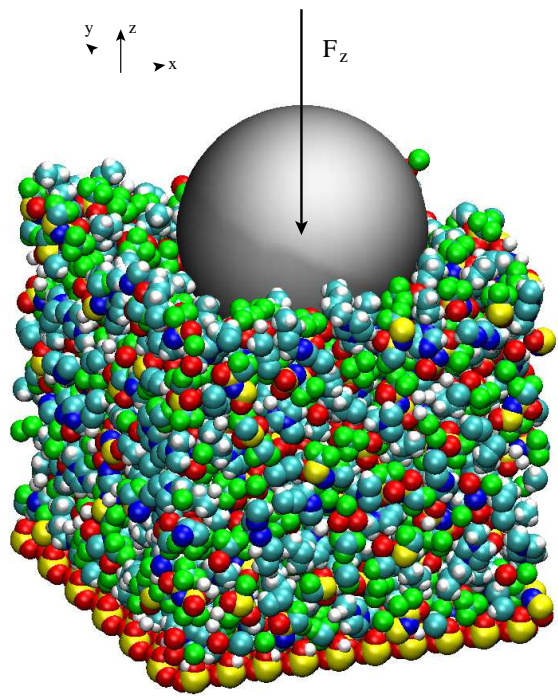}
\caption{Schematic view of the simulated system, showing the orientation of the reference axes.
}
\label{scheme}
\end{figure}

A schematic view of the system, showing the orientation of the reference axes, is given in Fig.~\ref{scheme}. Each simulation of an indentation 
event starts with the NS far above the free RTIL surface. The NS-surface distance is reduced, until when a sizable interaction is felt at the 
probing tip. At first the interaction is attractive, pulling the tip towards the liquid. The force becomes repulsive at $Z\sim 6$~nm, roughly 
equal to the film thickness plus the NS radius, marking the beginning of the {\it contact} range for our simulated AFM measurement. From this 
point of first contact, the tip is lowered towards silica through the RTIL in discontinuous steps of regular amplitude. At each distance, the 
system configuration is relaxed and then statistics is accumulated keeping fixed the relative position of the tip with respect to the silica 
plane. The primary result of our simulations is the average force $F_z$ on the meso-particle along the direction $z$ perpendicular to the silica surface. 

\begin{figure}[tbp]
\includegraphics[scale=0.6]{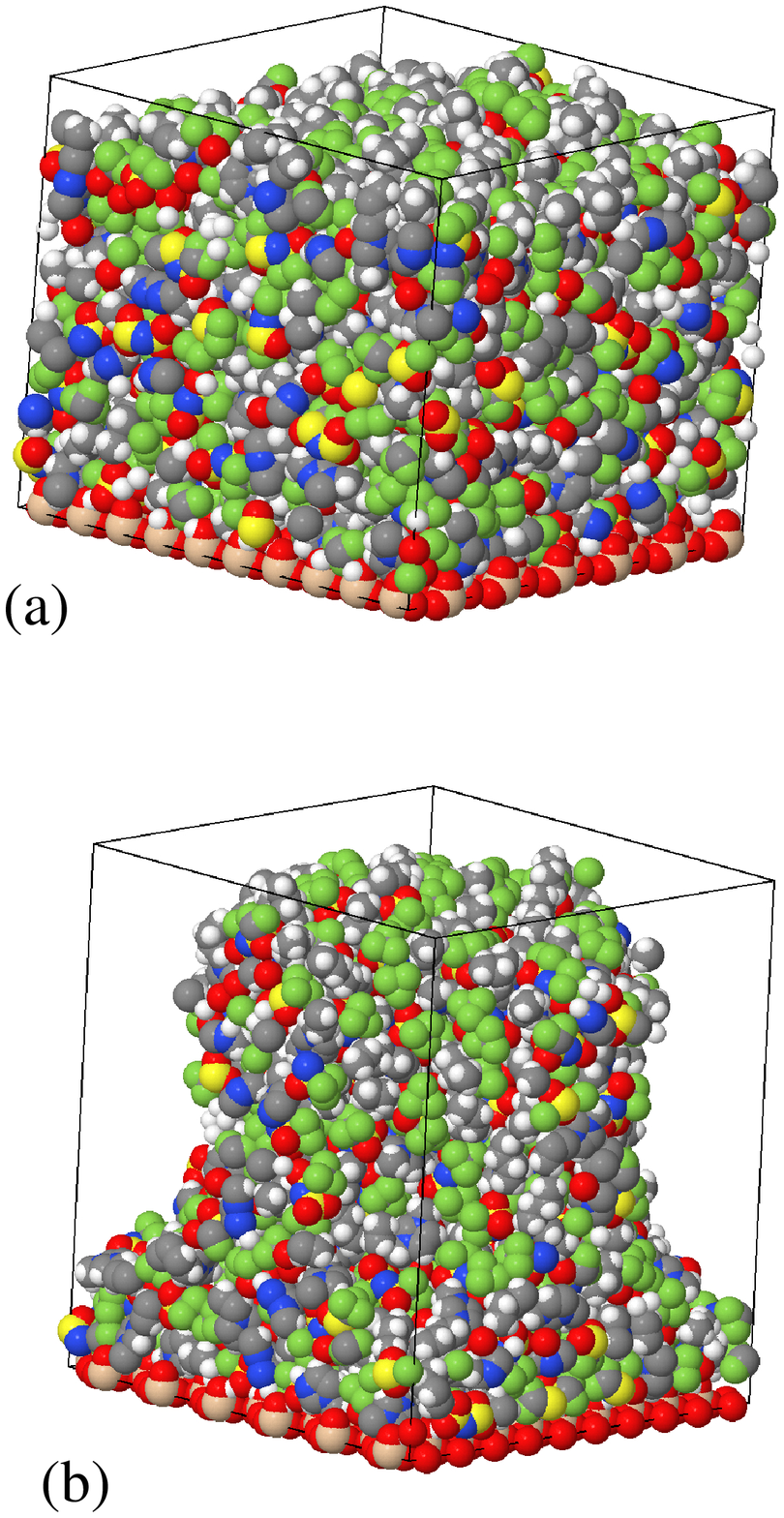}
\caption{
Two simulation snapshots, (a) before  and (b) during the indentation process.
The origin of the simulation box has been displaced by $( L_x/2; L_y/2, 0)$ with respect to
Fig.~\ref{scheme} to show the hole produced by the incoming nanoparticle.
Periodic boundary conditions are applied.
}
\label{snap}
\end{figure}

Simulations have been carried out at two temperatures, i.e., $T=300$~K and $T=350$~K. Four trajectories have been generated at $T=300$~K simulating 
the indentation by a particle of $\sigma=1.6$~nm. The relative size of the $\sigma=1.6$~nm NS and of the surface area can be appreciated in 
Fig.~\ref{scheme} and in Fig.~\ref{snap}. 
Independent starting configurations for the four distinct simulations were obtained by equilibrating at $T=300$~K four configurations
selected at 1~ns time separation along an equilibrium run at $T=350$~K, temperature at which we do observe diffusion of the RTIL ions.

\begin{figure}[tbp]
\includegraphics[scale=0.6]{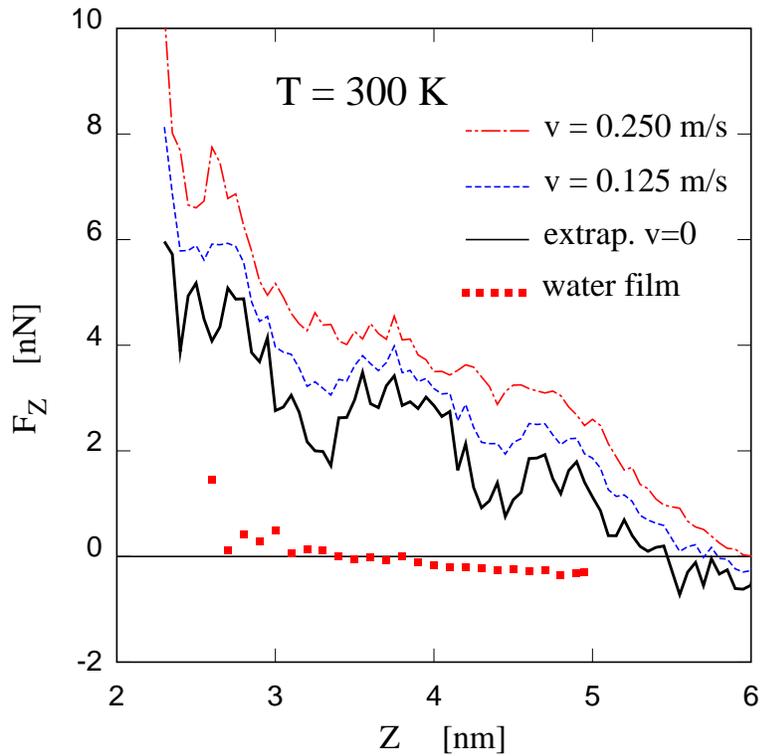}
\caption{
Average force on the incoming sphere as a function of distance from the silica surface. Dot-dashed line: 
approach velocity $v=0.25$~m/s; dashed line: approach velocity $0.125$~m/s; full line: extrapolation to $v=0$ 
(see text). Filled squares: force required to indent a water film of comparable thickness; single trajectory, $v=0.25$~m/s, $T=300$~K.
}
\label{averfz}
\end{figure}

For each trajectory, the $2.1 \leq Z \leq 6.0$~nm range has been covered in steps $0.05$~nm wide, relaxing the system at each distance
during $0.1$~ns, and collecting statistics during another $0.1$~ns simulation. Each simulation of the film indentation covers 
$16$~ns, and requires $\sim 10$ days running in parallel on 8 octuple-core Xeon E5530 2.4~GHz nodes. 
The average force $F_z$ as a function of distance $Z$ estimated during the simulation at $T=300$~K, $\sigma=1.6$~nm, is shown in Fig.~\ref{averfz}. 
The data refer to the average over time and over the four simulated trajectories. The average force is repulsive for $Z \leq 5.9$~nm, and its 
modulus tends to increase with decreasing separation, reaching a region of very high stiffness for $Z \leq 2.2$~nm. 
Even at the lowest separation and highest force, there remains one residual layer in between the tip and the surface.  

The $F_z$ dependence on $Z$ is not monotonic, but displays both localised irregularities and longer-wavelength small-amplitude oscillations. 
As discussed below, the longer wavelength oscillations reflect the defective but still apparent layering at the interface, while the irregularities
are due to activated molecular displacements representing rare events on the rather short time scale of our simulations.
In principle, the role of these last events could be reduced at will by increasing the number of trajectories over which $F_z$ is averaged, or
by increasing the time scale of the simulation.

To provide a comparison, we simulate the penetration of a 2~nm water film on silica, using the same NS and the same 
time progression. The results, represented by squares in Fig.~\ref{averfz}, show that the response of [bmim][Tf$_2$N] is, as expected, 
qualitatively different from that of water over most of the film thickness. Only at the shortest separations, do the computed forces 
become comparable, in both cases determined by the NS contact with a residual molecular layer pinned at the silica surface.

The average approach velocity in our simulations is $v=0.25$~m/s, that, although modest on a macroscopic scale, it is still
orders of magnitude faster than in AFM experiments, whether in tapping mode, and, even more, in force-distance mode. More importantly, the 
simulation time is still relatively short compared to the long relaxation times estimated in Ref.~\onlinecite{us} for [bmim][Tf$_2$N] 
at room temperature.
The $F_z(Z)$ force measured in our simulation, in particular, is expected to contain a friction contribution $F_z^f(Z)$, proportional to the 
approach velocity $v$, plus an equilibrium contribution, accounting for the $Z$-dependence of the system free energy. 
To compare our data with experiments, therefore, it is important to estimate and subtract the friction contribution $F_z^f(Z)$ to $F_z(Z)$. 
To this aim, we repeated the simulation at half the approaching speed, increasing from $0.1+0.1$~ns to $0.2+0.2$~ns the equilibration and averaging
time spent at each distance. The total simulation time is now $32$~ns per trajectory, and, because of the fairly high number of atoms, 
and of the long-range interactions, it represents a sizable computation. The new $F_z(Z)$, resulting 
from averaging over two independent trajectories, is shown in Fig.~\ref{averfz}. 
Since the friction contribution is proportional to $v$, the difference between the two curves
at $v$ and at $v/2$ represents $50$\% of $F_z^f(Z)$ measured at $v$, or $100$\% of $F_z^f(Z)$ measured at $v/2$. These simple considerations 
allow us to extrapolate our data to zero velocity, see the full line in Fig.~\ref{averfz}.
The relatively short time scale of our simulations does not allow us to exclude that highly activated events
corresponding to the migration of ions from underneath the tip, rarely seen in our simulations, and thus not fully accounted for in our extrapolation, 
could reduce the average force below the limit estimated by the relaxation on a short time scale. However, simulations at intermediate
values of $Z$ extended for a few ns beyond the standard relaxation time provided an estimate of the long-time value of $F_z$ consistent
with the extrapolation, enhancing our confidence in the results.

\begin{figure}[tbp]
\includegraphics[scale=0.7]{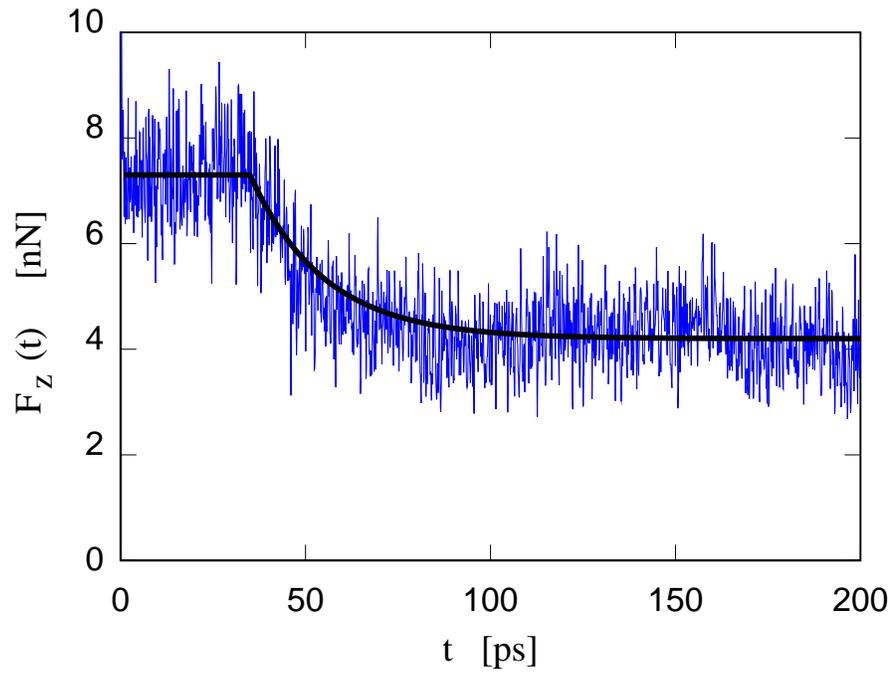}
\caption{
Time dependence of the force on the tip following the discontinuous displacement at $t=0$ of the incoming NS
from $Z=2.7$~nm \ to $Z=2.65$~nm. The thick line is a guide to the eye.
}
\label{timevol}
\end{figure}

The in-plane forces ($F_x$, $F_y$) acting on the tip average to zero because of symmetry. 
The average of their fluctuating modulus increases monotonically from $6 \geq Z \geq 4$~nm, reflecting the
increasing contact area between the incoming sphere and [bmim][Tf$_2$N].
The contact area stops increasing for $Z \lesssim 4$~nm, and also
the average modulus of $F_x$, $F_y$ remains nearly constant. A similar behaviour is displayed by the fluctuations of $F_z$ around its average $Z$-dependent value.
More interesting is the evolution of the instantaneous value of $F_z$ following a discontinuous decrease of $Z$ by $0.05$~nm, showing
the superposition of rapid thermal fluctuations with a slow downwards drift. A few discrete processes followed by exponential relaxation are
also seen from time to time, as exemplified in Fig.~\ref{timevol}.
Analysis of simulation snapshots shows that these jumps in the force are due to the large-amplitude motion of single ions, being displaced by
the increased pressure under the tip. These displacements represent rare activated events, and their random occurrence is likely to be the main 
cause of the short-length noise seen in the $Z$ dependence of $F_z$. 

\begin{figure}[tbp]
\includegraphics[scale=0.7]{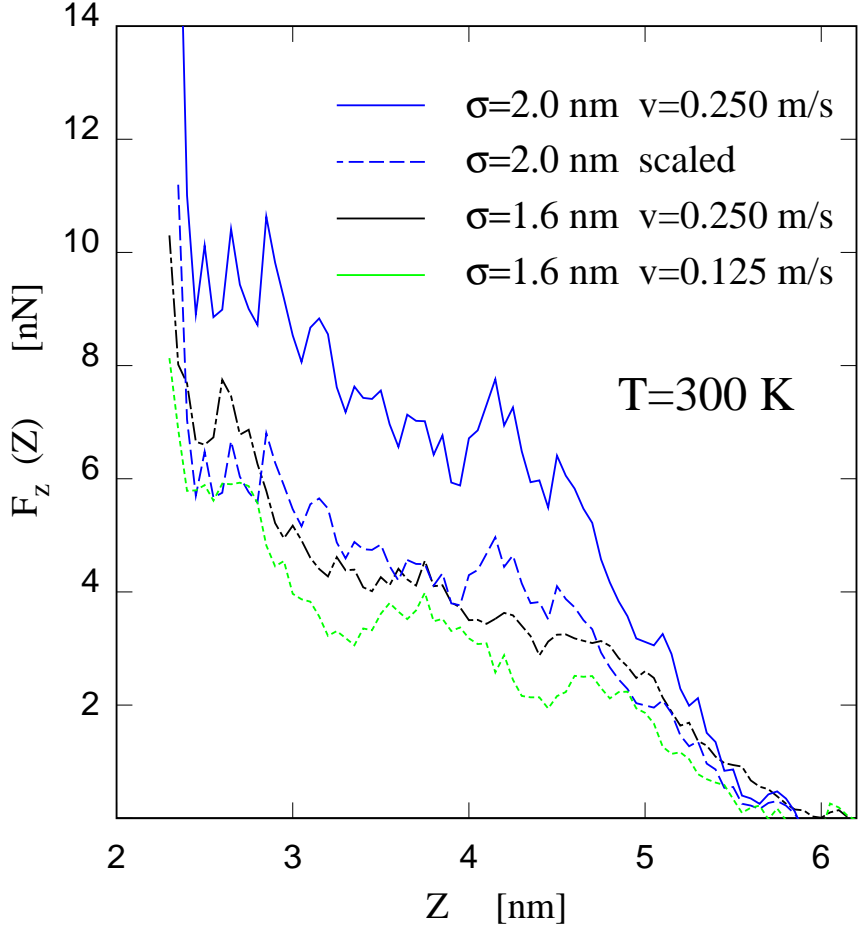}
\caption{Verification of the $F_z$ scaling with the tip radius $\sigma$. 
Full line, blue: average force on the $\sigma=2.0$~nm tip, shifted by $\Delta Z=-0.4$~nm.
Single trajectory at $v=0.125$~m/s, $T=300$~K. 
Dashed line, blue: average force on the $\sigma=2.0$~nm tip, shifted by $\Delta Z=-0.4$~nm and
multiplied by the ratio $(1.6/2.0)^2$ to highlight the scaling of $F_z$ with the contact area.
Green dotted line: average force on the $\sigma=1.6$~nm tip, single trajectory at $v=0.125$ m/s, $T=300$ K.
Black dash-dotted line: average force on the $\sigma=1.6$~nm tip, averaged over four trajectories at $v=0.25$~m/s, $T=300$~K.
}
\label{sigma2}
\end{figure}

The scaling of the force $F_z$ with the tip radius $\sigma$ is expected to be $F_z \propto \sigma^2$. To verify this relation, we
carried out simulations with a penetrating sphere of $\sigma=2$~nm. To account for the longer time needed to relax a larger perturbed volume,
the indentation has been carried out at the slowest speed $0.125$~m/s, and, because of cost considerations, only one trajectory has been
generated in this case. As expected, the force on the tip is systematically higher than in the simulations with $\sigma=1.6$ nm, as can 
be verified in Fig.~\ref{sigma2}), where the results for $\sigma=2$ nm are compared to those for $\sigma=1.6$~nm at $v=0.125$ m/s and 
at $v=0.250$ m/s.  Multiplication of the $\sigma=2$~nm results by $(1.6/2.0)^2$ brings them in between the two curves obtained with the 
$\sigma=1.6$~nm tip, thus confirming the expected size scaling.
On average, the rescaled data for $\sigma=2$~nm, $v=0.125$ m/s are somewhat closer to the $\sigma=1.6$~nm results obtained at the higher 
velocity $v=0.250$ m/s, probably because of the larger residual effect of relaxation for the $\sigma=2$~nm probe with respect to the
$\sigma=1.6$~nm case.

The observation that the average force scales with the expected $F_z \propto \sigma^2$ law provides an indirect test of convergence
with respect to the lateral periodicity $L_x$ and $L_y$. Spurious effects due to the finite size of the simulation cell have no reason 
to follow the same $\propto \sigma^2$ relation. The verification of $F_z \propto \sigma^2$, therefore, indirectly confirms that the finite-size 
error on $F_z$ is relatively small.

\begin{figure}[tbp]
\includegraphics[scale=0.7]{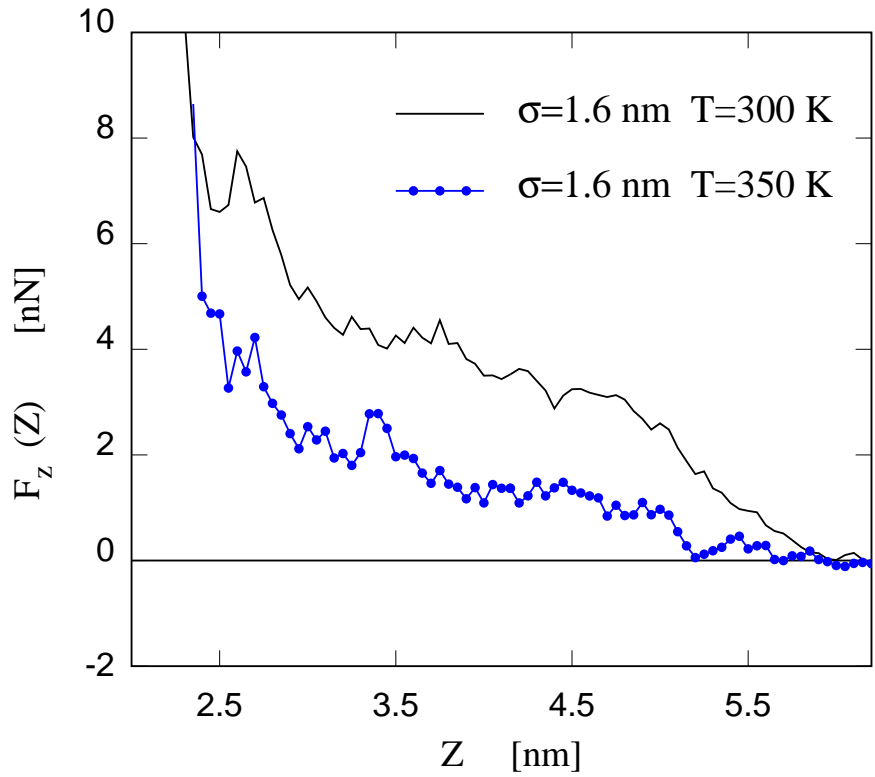}
\caption{
Full line: Average force on the $\sigma=1.6$~nm tip, averaged over four trajectories at $v=0.25$~m/s, $T=300$~K.
Full line with dots (blue): average force on the $\sigma=1.6$~nm tip, single trajectory at $v=0.25$~m/s, $T=350$~K.
}
\label{fzofT}
\end{figure}

To assess the effect of temperature, we carried out a simulation using a sample equilibrated at $T=350$~K. No extrapolation of $F_z(Z)$ to $v=0$ 
has been done in this case. A comparison with the raw data at $T=300$~K at the same approach velocity $v=0.25$ m/s shows that the film resistance 
to penetration decreases rapidly with increasing $T$ (see Fig.~\ref{fzofT}). At $T=350$~K, however, it is still much higher than the force estimated 
at $T=300$~K for the most prototypical liquid of comparable melting point, i.e., water.

As a final point, we verified that at $T=300$~K (but not at $T=350$~K) the deformation of the RTIL film due to the indentation is permanent on the 
simulation time scale, since the hole made by the approaching sphere
remains practically unchanged during several ns after removing the NS
potential, providing further evidence of the glassy state of the thin film.
Moreover, at $T=300$~K, the MD simulation of retracting the sphere, reversing the regular progression described above, invariably gives rise to small 
and negative forces, pointing to a large hysteresis in $F_z$. This effect is much attenuated at $T=350$~K.

\section{Comparison with experimental results}
\label{expe}

The conditions of our simulations are close to those of the measurements of Ref.~\onlinecite{milani1} on thin [bmim][Tf$_2$N] films deposited on 
the silica surface, while they differ somewhat from those of Ref.~\onlinecite{atkin}, carried out for different ionic liquids, and for thick, 
bulk-like RTIL films.

At first sight, despite several qualitative similarities, the simulation results seem to differ from both sets of experimental data. However,
as discussed below, new AFM results reconcile the simulation picture with the results of Ref.~\onlinecite{milani1}, and explain the
discrepancy with the $F_z$ measurements of Ref.~\onlinecite{atkin}.

We first discuss the comparison with the results of Ref.~\onlinecite{milani1}. First of all, the pressure under the tip required to penetrate the 
RTIL film is estimated by the simulation at $\sim 5$ kbar. This value is likely to be much higher than the maximum pressure applied during tapping 
mode measurements, estimated at $\sim 0.2$ kbar.  This implies that the AFM tip operated in tapping mode is unable to
penetrate the film, which therefore appears solid-like. 
On the other hand, the simulation results do not provide any explanation of the regular $0.6$~nm periodicity seen in the experiments. 
In the simulation sample, layering is apparently very defective, possibly as a result of the very long relaxation times of [bmim][Tf$_2$N] 
in close contact with the silica surface. Moreover, the $0.6$~nm periodicity seen under tapping mode conditions ($\sim 0.2$ kbar pressure under the tip)
might not be present at the pressures required to penetrate the film. Finally, RTIL are known to give rise to mesophases\cite{meso, perkin1, perkin2}, 
and a minimum system size larger than our simulated sample might be needed to reproduce the experiments quantitatively.

Comparison of the simulation results with the data of Ref.~\onlinecite{atkin} also shows similarities and differences. In particular, the 
comparison of our Fig.~\ref{averfz} with Fig.~\ref{timevol} of Ref.~\onlinecite{atkin} (a) shows similar qualitative features in the $Z$-dependence of 
computed and measured $F_z$, both displaying an increasingly repulsive character with decreasing $Z$, superimposed to localised fluctuations, and to
low-amplitude, long wavelength oscillations. Moreover, the intensity of the normal force $F_z$ is comparable in the two sets of data, in both cases
raising up to $\sim 16$ nN. This good agreement, however, is accidental, since these forces are measured on tips of significantly different radius, 
i.e., $2$~nm for the simulation, and $20$~nm for the experiment. The pressure under the tip, therefore, is significantly larger in simulation
than in experiment. This can be due, first of all, to the different ionic liquid considered in the simulation and in the experiment. 
More importantly, however, as already emphasised, the simulation refers to a thin film, while the measurements of Ref.~\onlinecite{atkin}
have been carried out with the AFM tip immersed into a thick bulk-like film. 

The crucial role of this last difference is confirmed by the results of a new set of AFM measurements\cite{milani2}, carried out by the same
team of Ref.~\onlinecite{milani1}.
Data for the normal force $F_z(z)$ measured by AFM
on a [bmim][Tf$_2$N] thin film deposited on
silica are shown in Fig.~\ref{expenew}. At variance from the conditions of Ref.~\onlinecite{milani1}, the force on the tip
is sufficiently large to penetrate the [bmim][Tf$_2$N] layers. These experimental results, therefore, refer to the same system that has been 
considered in the simulations, at comparable thermodynamical and probing conditions. Encouragingly, the data of Ref.~\onlinecite{milani2}
are the closest to the computational results.

More in detail, the results for $F_z$ in Fig.~\ref{expenew} show a few broad peaks of width $\sim 2$~nm, possibly 
corresponding to the tip penetration into successive molecular or ionic layers. In the case of a $10$~nm [bmim][Tf$_2$N] overlayer,
four successive rupture events can be clearly recognised, and some other sub-structures could be hidden by noise.
The indentation behaviour of [bmim][Tf$_2$N] islands is very reproducible, and curves acquired on different islands of similar height tend to 
overlap, including sub-structures. Remarkably, under these experimental conditions no $0.6$~nm regularity is apparent, even though 
the analysis of the topographic data of Ref.~\onlinecite{milani1} suggests that solid-like islands consists of stacks of layers $0.6$ nm thick.

A simple geometric estimate of the contact area shows that the peaks in the $F_z(z)$ of Fig.~\ref{expenew} correspond to a critical
pressure of about $3.5$ kbar, fully comparable to the $5$ kbar estimated for the simulated system. The residual quantitative discrepancy
might be due to differences in the time and size scales probed by experiments and by simulation, far beyond the range that we
can approach in our computations.

On the other hand, the electrostatic charge on the surface and on the tip seems to have only a minor effect on the measured normal force,
at least at the pressures required to penetrate the nanometric film. This has been indirectly verified by repeating the indentation
simulation on the mica/[bmim][Tf$_2$N], taking into account the surface charge that spontaneously form at the mica surface\cite{daniele}.
The simulation results reveal quantitative changes with respect to the silica case, but no qualitatively new features.

As a further check, experimental measurements of the $F_z$ force have been carried out on a bulk-like slab of [bmim][Tf$_2$N] on the mica 
surface\cite{milani2}. The results, not shown here, do reproduce those of Ref.~\onlinecite{atkin}, and confirm, in particular, that local
mechanical properties at the interface are different for thin and thick [bmim][Tf$_2$N] layers on a solid support. 

\section{Summary and conclusive remarks}
\label{conclusions}

Our results can be summarised as follows. We have simulated the indentation of a [bmim][Tf$_2$N] thin film deposited on an ordered silica 
surface by a structureless sphere of nanometric diameter. MD simulations have been carried out at $T=300$~K and at $T=350$~K,
based on an atomistic empirical potential model. The main result of our computation is the determination of the average force $F_z$ on the 
tip as  function of distance $Z$ from the planar silica surface.
The effect of different (and relatively high) approaching velocity has been estimated, and used to extrapolate the simulation results to 
conditions close to the experimental ones. Systematic errors in the extrapolation of the simulation data to zero approach velocity cannot be excluded 
and might be due to non-linear viscoelastic effects. These, however, are unlikely to change qualitatively the results, and a limited series of test 
supports the reliability of our extrapolation.

The scaling with the tip radius $\sigma$ has been verified to match the expected $F_z \propto \sigma^2$.

We verified by additional simulations for the unperturbed slab (i.e., no nanosphere) that an extra charge on the silica surface does not affect 
much the periodicity and amplitude of the RTIL density oscillations (see also Ref.~\onlinecite{mica}). 
However, it might have a larger effect on the regularity and persistence of {\it charge} oscillations\cite{sha, us}. 
These, in turn, could give rise to regularly oscillating forces on a charged tip, extending up to fairly long distances from the silica surface. 
While too weak to affect the result of
AFM force-distance measurements, reaching pressures of the order of a few kbar, these oscillating forces might 
become relevant on the $\sim 0.2$ kbar pressure scale that is the norm for
topographic measurements in tapping
or contact
mode, and could possibly explain the
regular, solid-like layered features
seen in Ref.~\onlinecite{milani1}. 

\begin{figure}[tbp]
\includegraphics[scale=1.0]{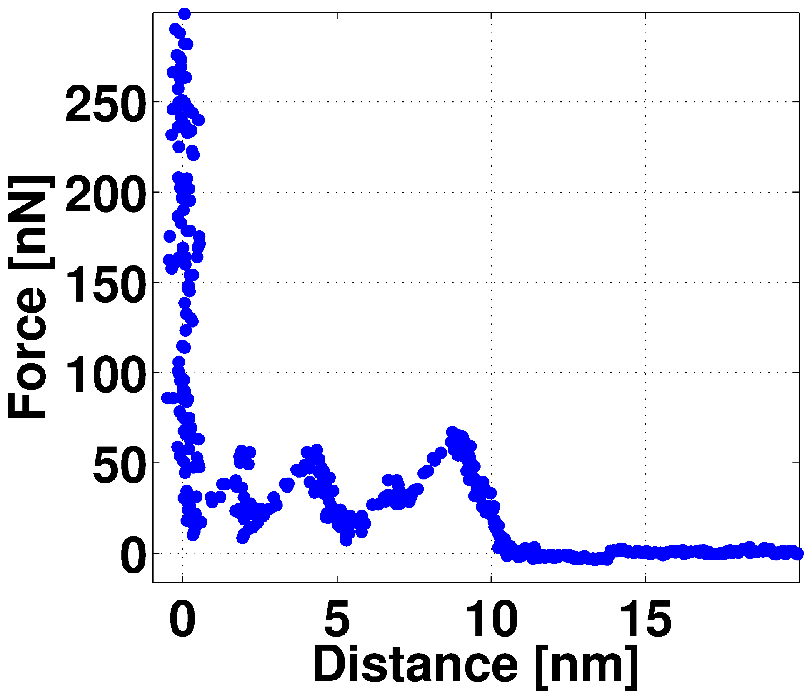}
\caption{
Normal force $F_z$ on a $\sim 20$~nm tip penetrating a $10$~nm [bmim][Tf$_2$N] film deposited on an oxidised silicon surface.
AFM force-distance measurements at room pressure and temperature\cite{milani2}. 
}
\label{expenew}
\end{figure}

Our results for the normal force $F_z$ differ from those reported in Ref.~\onlinecite{atkin}. However, the systems considered in 
these experiments and in our simulation differ for the choice of the ionic liquid, and for the conditions of the measurements.
Our simulations concern the indentation of a thin [bmim][Tf$_2$N] layer on silica, while the measurements of Ref.~\onlinecite{atkin}
have been carried out with the tip immersed into a bulk-like ionic liquid layer. The crucial difference between these two situations
is confirmed by new AFM measurements\cite{milani2}, whose results most
relevant for our study are being
anticipated in Fig.~\ref{expenew}.

Needless to say, quantitative differences remain between simulation and experimental results. Both experiments and simulations are
likely to be responsible for the discrepancy. For instance, the elasticity of the silica surface, neglected in our simulation, is likely 
to affect the force measured on the tip.
Differences in the response could also be due to a different quality and purity of the silica surface, including its hydroxylation coverage. 
Unfortunately, these parameters are difficult to control in experiment, and to include in simulation models. 
The technique to prepare and distribute droplets, relying on dissolving [bmim][Tf$_2$N] in methanol, and 
then letting the solvent to evaporate, might also play a role. 

In conclusion, we have shown that simulations can provide a nearly quantitative description of the indentation
of RTILs in contact with an insulating oxide surface, despite major technical limitations that still affect the computational approach\cite{extra}.
The resolution, time and length scales of the present simulations are not sufficient to reproduce and explain the solid-like features
seen in the topography of
thin [bmim][Tf$_2$N] films on oxide surfaces, but the gap between simulations and experiments is not very wide,
and could be reduced with only incremental changes of the simulation capability.
%

The observation of solid-like features in RTILs films on solid surfaces is, at the same time, conceptually intriguing and potentially very 
relevant for applications. Hints of similar phenomena in thin RTILs films have been reported a few times without a clearcut explanation.
Further evidence on these effects and phenomena has been provided by measurements carried out on RTILs confined into carbon\cite{carb} and
silica\cite{sili} matrices, whose porosity amplifies the role of RTIL/solid interfaces. Layering and slow relaxation times
might cause the solid-like properties of iono-gels made of oxide nanoparticles dissolved into RTILs\cite{graetzel}. 
Clarification of their stability and properties could greatly help their usage in electrochemistry and photovoltaic applications.

\vskip 3.0truecm
Acknowledgments -
The CIMaINa group has been financially supported by Fondazione Cariplo under the grant ``Materiali e tecnologie abilitanti 2007''.
The
collaboration has been made possible by an International Joint Project Grant of the Royal Society.

\begin{thebibliography}{99999999}
\bibitem{welton} T. Welton, {\it Chem. Rev.}, 1999, {\bf 99}, 2071. 
\bibitem{solvents} S. G. Cull, J. D. Holbrey, V. Vargas-Mora, K. R. Seddon, and G. J. Lye, {\it Biotechnol. Bioeng.}, 
2000, {\bf 69}, 227.
\bibitem{cath} V. I. Parvulescu and C. Hardacre, {\it Chem. Rev.}, 2007, {\bf 107}, 2615.
\bibitem{lubri} C. Ye, W. Lui, Y. Chen, and L. Yu, {\it Chem. Commun.}, 2001, 2244.
\bibitem{naina} J. J. Nainaparampil, K. C. Eapen, J. H. Sanders, and A. A. Voevodin,
{\it J. Microelectromech. Syst.}, 2007, {\bf 16}, 836.
\bibitem{electrochem} S. Zein El Abedinand, F. Endres, {\it Acc. Chem. Res. }, 2007, {\bf 40}, 1106; 
F. Endres, {\it ChemPhysChem}, 2002, {\bf 3}, 144; S. Zein El Abedinand, E. Mustafa, R. Hempelmann, H. Natter and F. Endres, 
{\it ChemPhysChem}, 2006, {\bf 7}, 1535.
\bibitem{photo} M. Gratzel, {\it Nature}, 2001, {\bf 414}, 338.
\bibitem{sum} J. B. Rollins, B. D. Fitchett and J. C. Conboy, {\it J. Phys. Chem. B}, 2007, {\bf 111}, 4990; C. R. Romero and S. Baldelli, 
{\it J. Phys. Chem. B}, 2006, {\bf 110}, 6213.  
\bibitem{xray} (a) E. Sloutskin, B. M. Ocko, L. Tamam, I. Kuzmenko, T. Gog, and M. Deutsch, {\it J. Am. Chem. Soc.}, 2005, {\bf 127}, 7796; 
(b) A. J. Carmichael, C. Hardacre, J. D. Holbrey, M. Nieuwenhuyzen, and K. R. Seddon, {\it Mol. Phys.}, 2001, {\bf 99}, 795.
\bibitem{neutro} J. Bowers, M. C. Vergara-Gutierrez, and J. R. P. Webster, {\it Langmuir}, 2004, {\bf 20}, 309.
\bibitem{perkin1} S. Perkin, L. Crowhurst, H. Niedermeyer, T. Welton, A. M. Smith, N. N. Gosvami, {\it Chem. Commun.}, 2011, {\bf 47}, 6572.
\bibitem{perkin2} S. Perkin, T. Albrecht, and J. Klein, {\it Phys. Chem. Chem. Phys.}, 2010, {\bf 12}, 1243.
\bibitem{ueno} K. Ueno, M. Kasuya, M. Watanabe, M. Mizukami, and K. Kurihara, {\it Phys. Chem. Chem. Phys.}, 2010, {\bf 12}, 4066.
\bibitem{naina2} J. J. Nainaparampil, B. S. Phillips, K. C. Eapen, and J. S. Zabinsli, {\it Nanotechnology}, 2005, {\bf 16}, 2474.
\bibitem{graetzel} P. Wang, S. M. Zakeeruddin, P. Compte, I. Exnar, and M. Gr{\"a}tzel, {\it J. Am. Chem. Soc. }, 2003, {\bf 125}, 1166;
K. Ueno, S. Imaizumi, K. Hata, and M. Watanabe, {\it Langmuir}, 2009, {\bf 25}, 825.
\bibitem{atkin} (a) R. Hayes, G. G. Warr, and R. Atkin, {\it Phys. Chem. Chem. Phys.}, 2010, {\bf 12}, 1709; (b)
R. Atkin and G. G. Warr, {\it J. Phys. Chem. C}, 2007, {\bf 111}, 5162; D. Wakeham, R. Hayes, G. G. Warr, and R. Atkin,
{\it J. Phys. Chem. B}, 2009, {\bf 113}, 5961.
\bibitem{milani1} S. Bovio, A. Podest{\`a}, C. Lenardi, and P. Milani, {\it J. Phys. Chem. B}, 2009, {\bf 113}, 6600.
\bibitem{milani2} S. Bovio, A. Podest{\`a}, and P. Milani, {\it unpublished}.
\bibitem{others} Y.-D. Liu, Y. Zhang, G.-Z. Wu, and J. Hu, {\it J. Am. Chem. Soc.}, 2006, {\bf 128}, 7456.
\bibitem{wippf} N. Seifert and G. Wipff,  {\it J. Phys. Chem. C}, 2008, {\bf 112}, 19590.
\bibitem{us} S. Bovio, A. Podest{\`a}, P. Milani, P. Ballone, and M. G. Del P{\'o}polo, 
{\it J. Phys.: Condens. Matter}, 2009, {\bf 21}, 424118. 
\bibitem{ruth} L. D. Gelb and R. M. Lynden-Bell, {\it Chem. Phys. Lett.}, 1993, {\bf 211}, 328;
L. D. Gelb and R. M. Lynden-Bell, {\it Phys. Rev. B}, 1994, {\bf 49}, 2058; 
D. L. Patrick and R. M. Lynden-Bell, {\it Surf. Sci.}, 1997, {\bf 380}, 224.
\bibitem{pote} J. N. Canongia Lopes, J. Deschamps, and A. A. H. Padua,
{\it J. Phys. Chem. B}, 2004, {\bf 108}, 2038;  J. N. Canongia Lopes and A. A. H. Padua,
{\it J. Phys. Chem. B}, 2004, {\bf 108}, 16893. See also (Additions and corrections):
J. N. Canongia Lopes, J. Deschamps, A. A. H. Padua, {\it J. Phys. Chem. B }, 2004,
{\bf 108}, 11250.
\bibitem{daan} P. R. ten Wolde and D. Frenkel, {\it Science}, 1997, {\bf 277}, 1975; M. H. J. Hagen and D. Frenkel, {\it J. Chem. Phys.}, 1994, {\bf 101}, 4093.
\bibitem{DLPOLY} W. Smith, M. Leslie, and T. R. Forester, DL$-$POLY v.2.14;
Daresbury Laboratories: Daresbury, Warrington, WA4 4AD, UK, 2003.
\bibitem{meso}  J. N. A. Canongia Lopes, A. A. H. Padua,  {\it J. Phys. Chem. B}, 2006, {\bf 110}, 3330; A. Triolo, O. Russina, H.-J. Bleif, 
E. Di Cola, {\it J. Phys. Chem. B}, 2007, {\bf 111}, 4641; N. Manini, M. Cesaratto, M. G. Del P{\'o}polo, and P. Ballone, {\it J. Phys. Chem. B},
2009, {\bf 113}, 15602.
\bibitem{daniele} Daniele Dragoni, Master Thesis, University of Milano, 2011. Available on-line at www.mi.infm.it/manini/theses/dragoni.pdf
\bibitem{mica} The same conclusion is supported by the simulation of a tip approaching the [bmim][Tf$_2$N]/mica interface, see Ref.~\onlinecite{daniele}.
\bibitem{sha} M.-L. Sha, G.-Z. Wu, Q. Dou, Z.-F. Tang, and H.-P. Fang, {\it Langmuir}, 2010, {\bf 26}, 12667.
\bibitem{extra} J. Tamayo and R. Garcia, {\it Langmuir}, 1996, {\bf 12}, 4430; R. Garcia and R. Perez, {\it Surf. Sci. Reports}, 2002, {\bf 47}, 197.
\bibitem{carb} J. Monk, R. Singh, and F. R. Hung, {\it J. Phys. Chem.}, 2011, {\bf 115}, 3034.
\bibitem{sili} B. Coasne, L. Viau, and A. Vioux, {\it J. Phys. Chem. Lett.}, 2011, {\bf 2}, 1150.
\end{thebibliography}
\end{document}